     \documentclass[12pt,preprint]{aastex}

      \usepackage{epsfig}

\newcommand{\ea}{et al.}

\newcommand{\s}{\>{\rm s}}
\newcommand{\kms}{\>{\rm km}\,{\rm s}^{-1}}
\newcommand{\pc}{\>{\rm pc}}
\newcommand{\kpc}{\>{\rm kpc}}
\newcommand{\mpc}{\>{\rm Mpc}}

\newcommand{\myr}{\>{\rm Myr}}
\newcommand{\gyr}{\>{\rm Gyr}}
\newcommand{\msun}{\>{\rm M_{\odot}}}
\newcommand{\lsun}{\>{\rm L_{\odot}}}

\newcommand{\as}{^{\prime\prime}}

\newcommand{\bdm}{\begin{displaymath}}
\newcommand{\edm}{\end{displaymath}}
\newcommand{\beq}{\begin{equation}}
\newcommand{\eeq}{\end{equation}}
\newcommand{\bit}{\begin{itemize}}
\newcommand{\eit}{\end{itemize}}
\newcommand{\ben}{\begin{enumerate}}
\newcommand{\een}{\end{enumerate}}
\newcommand{\bfi}{\begin{figure}[htb]}
\newcommand{\bpfi}{\begin{figure}[p]}

\newcommand{\nc}{nuclear cluster}

\slugcomment{to appear in AJ}

\shorttitle{Census of Nuclear Star Clusters}
\shortauthors{B\"oker \ea }

\begin{document}





\def\xx{$^{[xx]}$}

\def\tbd#1{{\baselineskip=9pt\medskip\hrule{\small\tt #1}
\smallskip\hrule\medskip}}

\def\com#1{{\baselineskip=9pt\medskip\hrule{\small\sl #1}
\smallskip\hrule\medskip}}

\title{An {\it HST} Census of Nuclear Star Clusters in Late-Type Spiral Galaxies: 
I. Observations and Image Analysis\altaffilmark{1}}


\author{Torsten B\"oker\altaffilmark{2}, Seppo Laine, Roeland P. van der Marel}
\affil{Space Telescope Science Institute, 3700 San Martin Drive, 
Baltimore, MD 21218, U.S.A.}
\email{boeker@stsci.edu, laine@stsci.edu marel@stsci.edu}

\author{Marc Sarzi, Hans-Walter Rix}
\affil{Max-Planck-Institut f\"ur Astronomie, K\"onigsstuhl 17, 
D-69117 Heidelberg, Germany }
\email{sarzi@mpia-hd.mpg.de, rix@mpia-hd.mpg.de}

\author{Luis C. Ho}
\affil{Observatories of the Carnegie Institution of Washington, 
        813 Santa Barbara Street, Pasadena, CA 91101-1292, U.S.A.}
\email{lho@ociw.edu}

\author{Joseph C. Shields}
\affil{Ohio University, Department of Physics and Astronomy, 
        Clippinger Research Laboratories, 251B, Athens, OH 45701-2979}
\email{shields@helios.phy.ohiou.edu}


\altaffiltext{1}{Based on observations made with the NASA/ESA 
{\it Hubble Space Telescope}, obtained at the Space Telescope Science 
Institute, which is operated by the Association of Universities for 
Research in Astronomy, Inc., under NASA contract NAS 5-26555. These
observations are associated with proposal \#\,8599.}
\altaffiltext{2}{Affiliated with the Astrophysics Division, Space
Science Department, European Space Agency}


\begin{abstract}
We present new {\it HST} I-band images of a sample of 77 nearby, 
late-type spiral galaxies with low inclination. The main
purpose of this catalog is to study the frequency and properties of
nuclear star clusters. In 59 galaxies of our sample, we have identified
a distinct, compact (but resolved), and dominant 
source at or very close to their photocenter. In many cases,
these clusters are the only prominent source within a few 
kpc from the galaxy nucleus.
We present surface 
brightness profiles, derived from elliptical isophote fits,
of all galaxies for which the fit was successful. 
We use the fitted isophotes at radii larger than $2\as$ 
to check whether the location of the cluster coincides with the 
photocenter of the galaxy, and confirm that in nearly all cases, 
we are truly dealing with ``nuclear'' star clusters. 
From analytical fits to the surface brightness 
profiles, we derive the cluster luminosities after subtraction of the 
light contribution from the underlying galaxy disk and/or bulge.
\end{abstract}
\section{Introduction}
Over the past decade, high dynamic range observations with 
modern CCD detectors have shown that compact stellar nuclei are a 
common feature of spiral galaxies of all Hubble types. For example,
\cite{mat97} found 10 objects with compact nuclear star clusters
in a survey of 49 southern, very late-type spirals. However,
as one progresses along the Hubble sequence 
towards earlier types, the increasingly luminous bulge
component with its steeply rising surface brightness profile 
makes the identification of an additional, unresolved cluster
extremely difficult.
It therefore took the unique spatial resolution of the {\it Hubble Space 
Telescope (HST)} to demonstrate that \nc s are a common
phenomenon also in earlier Hubble types \citep*[e.g.][]{car98}. 
The {\it HST} currently
provides the only means to investigate the structural properties
of nuclear star clusters, as demonstrated by \cite{mat99}, and to
cleanly separate their emission from the underlying galaxy disk/bulge.

Despite the recent progress, the formation mechanism of 
nuclear star clusters remains largely a mystery. Intuitively, 
there are good reasons to expect matter accumulation
in the deep gravitational wells of galaxies with massive bulges, and
hence active star formation in their nuclei.
In contrast, the gravitational force all but vanishes in the
centers of pure disk galaxies with shallow surface brightness
profiles and without any discernible bulge component. In these
galaxies, the dynamical center is not a ``special'' place and it is
far from obvious how a massive stellar cluster could have formed there.
The shallow gravitational potential might provide a natural explanation for
the fact that spirals of late Hubble type are not known to contain
super-massive black holes. On the other hand, nuclear star clusters 
can be extremely compact: the nucleus of M33, for example, has
likely undergone core collapse and is as compact as
any known globular cluster \citep{kor93}. So far, no satisfying 
explanation has been put forward to explain the high gas densities 
that must have been present in the nuclei of these shallow disk galaxies
to enable the formation of such massive and compact objects.

It is also unknown whether nuclear star clusters form repeatedly
or only once - a question with important implications for the 
dynamical and morphological evolution of their host galaxies. 
To make progress along this line, it is essential to obtain the age 
distribution of nuclear star clusters.
So far, reliable age estimates exist for only a handful of nuclear
star clusters. Interestingly, most of them appear to be rather young: 
our Galaxy has a central stellar cluster with an age of only 
$\sim 3$~Myrs \citep{kra95}, and both M31 and M33 have
blue nuclei that are very likely young star clusters \citep{lau98}. 
More recently, we have published \nc\ ages derived
from ground-based spectroscopy for IC~342 
\citep*[$\rm \leq 60\>$Myrs,][]{boe99}, and NGC~4449 
\citep*[$6-10\>$Myrs,][]{boe01,gel01}. 
In addition, the dominant stellar population of the nuclear 
cluster in NGC 3227 is less than 50~Myrs old \citep*{sch01}.

However, it is possible (and in fact quite likely) that ground-based 
observations predominantly target the brightest and hence youngest clusters. 
In order to get a more representative picture of nuclear star clusters,
it is important to study a galaxy sample which is free from selection
effects that favor the high end of the \nc\ luminosity range.
In this paper, we describe the results of an {\it HST} I-band imaging survey 
of an unbiased sample of nearby, face-on, very late-type spirals 
(Scd or later). The main goals of the survey are (a) to determine the 
frequency of nuclear star clusters in very late-type spirals, 
(b) to derive their luminosity and size distribution, 
(c) to compare their properties to those of nuclear star clusters
in earlier Hubble types which have been more extensively studied with {\it HST}
\citep{car97,car98,car01,car99}, and (d) to provide a source catalog for 
follow-up spectroscopic observations to age-date their stellar populations.
The main purpose of this paper is to present the complete dataset. In a
companion paper (B\"oker \ea\ 2002, in preparation), 
we describe the statistics
of the full sample and investigate whether the properties of nuclear star
clusters correlate in any way with those of their host galaxies. 
 
This paper is organized as follows: in \S~\ref{sec:obs}, we describe
our sample selection criteria, the observational strategy, and the
data reduction procedure, and we present the final images as well as the 
results of the isophotal analysis. In \S~\ref{sec:disc}, we discuss 
whether the clusters indeed occupy the nuclei of their host galaxies,
and how they compare to other luminous star clusters observed in a variety
of starburst environments. We conclude in \S~\ref{sec:concl}.
\section{Sample Selection, Observations and Data Reduction}\label{sec:obs}
\subsection{The Sample}\label{subsec:sample}
The target list for our survey was selected from the RC3 catalog of bright 
galaxies \citep{dev91} according to the following criteria:
\ben
\item{Hubble type between Scd and Sm ($\rm 6 \leq T \leq 9$).}
\item{Line-of-sight velocity $\rm v_{\rm hel} < 2000\kms$, 
to assure good spatial resolution in physical units.}
\item{Axis ratio parameter ${\rm R_{25}} \equiv log(a/b) < 0.2$, i.e. 
inclination close to face-on. This helps to minimize the effects of 
dust extinction due to the galaxy disk, and to avoid confusion in the
identification of the nucleus from line-of-sight projection of 
disk clusters.}
\een

Our sample is unbiased with respect to galaxy size, mass,
total magnitude, star formation efficiency, or any other quantity
that might reasonably be expected to favor or disfavor \nc\ formation.
It should therefore be well suited to provide an objective census
of \nc s in late-type galaxies in the local universe.

We identified a total of 113 galaxies which satisfied the above criteria
and had not previously been observed with the Wide Field and Planetary
Camera 2 (WFPC2) onboard {\it HST} in the F814W
filter. These 113 galaxies were used as the target pool for our
WFPC2 snapshot program (GO-8599). To date, 77 galaxies have been 
successfully observed, listed in Table~\ref{tbl:obs}. 
It is possible that a few more targets will be observed later, but
for this paper, we limited the sample to those galaxies observed before 
August 3, 2001.

\subsection{The WFPC2 Images}\label{subsec:images}
All images were taken with the WFPC2 camera onboard {\it HST}, with
the galaxy nucleus centered on the Planetary Camera (PC) chip. 
The PC pixel size is $0.046''$, and the field of view is 
$36'' \times 36''$. We 
used the F814W filter with an integration time of $600\s$,
split in two exposures of $300\s$ to allow cosmic ray rejection.
We also took a short exposure ($40\s$) to guard against saturation
of the WFPC2 detectors in the $300\s$ exposures. However, none of the 
galaxies was bright enough to require the use of the $40\s$ exposure.
The PSF with the F814W filter has a FWHM of $0\farcs 07$.

We used the STSDAS task {\tt wfixup} to interpolate (in the x-direction) over bad
pixels as identified in the data quality files. We also used the STSDAS task
{\tt warmpix} to correct consistently warm pixels in the data, using the
most recent warm pixel tables which are provided by the WFPC2 instrument
group at STScI about once a month. The STSDAS task {\tt ccrej} was used to 
combine the two $300\s$ exposures. This step 
corrects most of the pixels affected by cosmic rays in the combined image.
In general, a few cosmic rays remain uncorrected, mostly when the same
pixel was hit in both exposures. Also, a small number of hot pixels remain
uncorrected because they are not listed even in the most recent warm pixel
tables. We corrected these with the IRAF task {\tt cosmicrays}, setting the 
``threshold'' and ``fluxratio'' parameters to suitable values that were selected
by careful comparison of the images before and after correction to ensure
that only questionable pixels were replaced. The photometric calibration,
and conversion to Johnson I-band was performed according to \cite{hol95}.
We assumed a standard color of $\rm V-I = 1$ for the galaxies which 
translates into a zeropoint of 21.55 (note that the assumed color affects
the zeropoint only weakly: a color of $\rm V-I = 2$ would result in a 
zeropoint of 21.56). 

After visual inspection of the images, we divided the objects into two
groups. The first group contains those 59 objects for which a) the 
(photo-) center of the galaxy is reasonably well defined, and b) a prominent, isolated
point-like source can be identified close to it. These sources are
\nc\ candidates. In \S~\ref{subsec:nuc?}, we will discuss our criteria
for whether they are indeed occupying the photocenter or not.

The second group contains the remaining 18 galaxies which 
show no easily identifiable source close to the center. 
This does not necessarily mean that these galaxies do not harbor a \nc ; 
it merely indicates that we cannot identify one with
any kind of certainty from our data. 
Figures~\ref{fig:maps1} and \ref{fig:maps2} contain 
the images of the 59 (18) galaxies in group 1 (2).

Visual inspection of the images reveals a number of noteworthy points: 
\ben
\item{In most cases, the nuclear star cluster candidate 
is obvious in the images,
because it is the dominant source at or close to the photocenter
of the galaxy. It is often (but not always) the brightest source in the 
field, and in many cases the only cluster within a kpc from the
photocenter. 
}
\item{For the vast majority of the sample galaxies, the images
show no morphological evidence for a stellar bulge. While our sample 
was obviously selected to avoid bright stellar bulges, it is still
surprising that there appears to be no bulge at all in many
late-type spirals. 
The visual impression is confirmed by the surface brightness analysis
in \S~\ref{subsec:sbp}. Our dataset is uniquely suited for a 
detailed investigation of the structural properties of 
late-type spiral galaxies which is, however, beyond the 
scope of this paper. We defer a more detailed study of the 
disk surface brightness, the (lack of) evidence for stellar bulges, 
and possible correlations with \nc\ properties 
to a later paper (Stanek \ea\ 2002, in preparation). 
}
\item{In many cases (e.g. UGC\,3574, UGC\,5015, or NGC\,4411B), the
\nc\ is ``naked'': it forms a distinct entity that appears 
completely isolated within the disk. The cluster 
location does not appear to be a dynamically ``special'' place, 
because there are no spiral arms, dust lanes, or other signs of a 
kinematic center visible in the images. This is even true at the
smallest spatial scales as observed in the most nearby galaxies, such
as NGC\,300 or NGC\,7793. This confirms a notion by \cite{mat99} who
studied a sample of four extreme late-type spirals, also with WFPC2.
}
\item{In other cases (e.g. NGC\,853, NGC\,2139, or NGC\,4027), however, 
the cluster location seems to be the origin of spiral structure or 
prominent dust lanes, indicative of it being at the dynamical center
of the galaxy.
If the mechanisms that lead to such a morphology are in any way
connected to the presence of a \nc , it appears that they are 
a consequence rather than the prerequisite of \nc\ formation, because
it is difficult to imagine how a galaxy like NGC\,2139 can change its
structure back to the smooth and regular appearance of those with
``naked'' nuclear clusters. 
}
\een
\subsection{Isophotal Fits}\label{subsec:sbp}
We used the IRAF-task {\tt ellipse} to obtain surface brightness
profiles (SBP) over the PC field of view for all galaxies in our 
sample. For the galaxies in group 1, we
started the fitting process centered on the cluster with
a semi-major axis (SMA) length of 5 pixels ($0\farcs 23$). We varied
the SMA length logarithmically (with a 15\% stepsize), first going out to
a maximum SMA of 350 pixels ($\approx 16\as$), then inwards to
the sampling limit (SMA of 0.5 pixels or $0\farcs 023$). 
Throughout the fit,
the ellipse center, ellipticity, and position angle were allowed to
vary freely. By comparing the position of the peak surface brightness
(i.e. the position of the putative \nc )
to the center of the outer isophotes, we were able to decide whether the
cluster indeed occupies the photocenter of the galaxy. This is 
further discussed in \S~\ref{subsec:nuc?}. 

For a small number of galaxies, such an
unconstrained fit failed for a small range of radii, typically
because of a complex morphology, a shallow surface brightness 
gradient, a low signal-to-noise ratio, or any combination of these factors. 
In these cases, we performed two fits, one as described above, and going 
out as far as possible, and a second fit starting at a large radius, 
going inward as far as possible.
By combining the two fits, we were able to construct the SBP 
over most of the radial range, with data missing
for only a few radii. 
For another small group of galaxies, we were forced to increase
the spacing between isophotes (stepsize of 50\%)
to overcome the low signal-to-noise ratio.

The 18 galaxies in group 2 have no plausible
candidate for a \nc . These are mostly objects with very
low surface brightness and an ill-defined photocenter. In
these cases, we proceeded as follows: we first derived an estimate for 
the photocenter and ellipticity of the galaxy from the average 
of three isophotes at large radii (typically 200 pixels plus or
minus 15\%). The SBP was then derived from a second fit for which
the isophote center and ellipticity was fixed to the initial estimates.
This procedure worked for all but four galaxies, for which it was impossible
to obtain even an estimate for the photocenter. One of these (ESO\,510-59)
appears to be a merger pair, one (NGC\,6946) contains large amounts of
dust in the nucleus, and the other two (A\,1301-03, and IC\,4182) 
are too faint in our images to detect a meaningful SB gradient.

In Figures~\ref{fig:profiles} and~\ref{fig:badprofiles},
we show the resulting SBPs for all galaxies in group 1
and 2, respectively.
The presence of the \nc\ candidate is obvious by the sharp upturn
in the SBPs of Figure~\ref{fig:profiles}, typically at radii 
around $0\farcs 3$. 
Not surprisingly,
the clear upturn is absent in the profiles of the group 2 galaxies
in Figure~\ref{fig:badprofiles}. Their SBPs are in general noisier,
and in some cases even decrease in brightness towards the center, just
another manifestation of their shallow surface brightness gradients.

\subsection{Photometry of the Nuclear Clusters}\label{subsec:phot}

For the derivation of the luminosity of the {\nc}s it is useful to
have a parametrized fit to the SBP. For this, we used the form
\beq
I(r) = I_0\cdot (r/r_b)^{-\gamma}
\cdot (1+ (r/r_b)^{\alpha})^{\frac{\gamma - \beta}{\alpha}}
\cdot (1+ (r/r_c)^{\delta})^{\frac{\beta - \epsilon}{\delta}}  .
\label{e:sbfit}
\eeq
This is based on the so-called `nuker-law' parametrization
\citep{lau95,byu96}, which represents a broken power-law with an
inflection point at a radius $r_b$. We added an additional factor 
which allows for the possibility of a second inflection point at 
a radius $r_c$. The resulting equation was found to be sufficiently 
general for the purposes of this paper.

In general, the presence of a nuclear cluster causes a distinct 
upturn in the SBP at a certain radius $R_{\rm u}$. This radius was
identified by eye for each galaxy, the adopted values are listed in
Column~5 of Table~~\ref{tbl:res}. To estimate the cluster luminosity,
we started by fitting the parametrization~(\ref{e:sbfit}) to the data
inside $R_{\rm u}$ (dotted curves in Figure~\ref{fig:profiles}),
followed by integration over an aperture with radius $R_{\rm u}$. To
obtain an estimate for the nuclear cluster luminosity, one needs to
subtract from this the light contribution of the galaxy disk (and
possibly bulge) within $R_{\rm u}$. To this end, we considered
two models for the SBP of the underlying galaxy light within the PC 
field of view which are likely to bracket the true SBP of the galaxy.
For the first model, we assumed that the underlying galaxy has a constant
surface brightness inside $R_{\rm u}$ (dashed lines in
Figure~\ref{fig:profiles}). For the second model, we performed a fit of a
nuker-law to the data outside $R_{\rm u}$, and extrapolated that fit
to radii inside $R_{\rm u}$ (solid curves in Figure~\ref{fig:profiles}).

After subtraction of the integrated luminosity inside $R_{\rm u}$ 
of the two models for the underlying galaxy light, we
obtain two different estimates for the {\nc} luminosity. In
Table~\ref{tbl:res}, we list the mean of these two estimates and also
half their difference as a measure of the uncertainty. The latter
uncertainty indicates only the extent to which the cluster photometry
depends on the choice of underlying galaxy model. In general, this is
not the dominant source of error. The uncertainty due to the exact
choice of the aperture size $R_{\rm u}$ adds at least $0.1$ mag of
error to the {\nc} luminosity estimates.

In Figure~\ref{fig:hist}, we plot histograms of both apparent and
absolute cluster luminosity. The distribution has a FWHM
of about 4 magnitudes, with a median of $\rm M_I = -11.5$.
This is brighter than even the brightest
globular cluster in the Milky Way, but comparable to the bright
end of the cluster luminosity function in the NGC\,4038/39
merging system \citep{whi99} or the young super star clusters
in M\,82 \citep{oco95}. An absolute luminosity of $\rm M_I = -11.5$ 
corresponds to $1.6\cdot 10^6 \lsun$ in the I-band (because ${\rm M}_{I,\odot} =
4.02$). The associated mass depends on the unknown mass-to-light ratio
$M/L$. For reference, one can consider the case of a cluster formed in
an instantaneous burst with a \cite{sal55} initial mass function and
solar metallicity. For a young cluster with an age of $10\myr$ one
then has $M/L_I \approx 0.016$ and $M = 2.6\cdot 10^4 \msun$, whereas for an old
cluster with an age of $5 \gyr$ one has $M/L_I \approx 0.43$ and $M =
6.9\cdot 10^{5} \msun$ \citep{lei99}. Our ongoing spectroscopic program to
derive cluster ages for many of the sample galaxies promises to remove
this ambiguity.

Photometry of the off-nuclear clusters in the low surface brightness
disks of the group 2 galaxies, e.g.~in UGC\,12082 or ESO\,187-51,
shows that we can easily detect clusters as faint as $\rm M_I = -8$. 
However, none of the galaxies in group 2 shows any evidence for
{\it nuclear} clusters in this luminosity range. This demonstrates
that the low-luminosity cutoff in Figure~\ref{fig:hist}b around $\rm
M_I = -9$ is probably real.

\subsection{The size of nuclear star clusters}
To derive physically meaningful information about the 
structural properties of \nc s such as the half-light radius 
or the core radius in a King model, the observed SBPs have to 
be corrected for the instrumental point spread function (PSF).
Since the clusters in most cases are not much 
more extended than the {\it HST} PSF, and the shape of the PSF
is rather complex because of its extended wings, the
deconvolution is a non-trivial task which we defer
to a later paper (Sarzi \ea\ 2002, in preparation).

For now, we list in Table~\ref{tbl:res} a simple
measure of the \nc\ sizes, namely the half-width-at-half-maximum (HWHM), 
i.e. the radius at which the {\it observed} surface brightness drops
to half its peak value. These were derived using simple
linear interpolation between the two datapoints in the SBP that
bracket half the peak value of the surface brightness.
The listed values can be compared to those for the
{\it HST} PSF. We constructed a PSF model from the TinyTim
software package \citep{kri01} for the PC chip and the F814W
filter, and performed an identical isophotal fit. The resulting 
HWHM was $0\farcs 032$. For comparison, an identical analysis
for a bright star in the image of the galaxy NGC\,6509 yields
$0\farcs 036$. Both values are smaller than those
listed in Table~\ref{tbl:res}, which confirms that the \nc s are
indeed resolved.  

Figure~\ref{fig:size_hist}a contains a histogram of the angular
HWHM distribution which is strongly peaked around $0\farcs 06$. 
We caution that the complexity of the {\it HST} PSF (which is
only poorly represented by a single Gaussian)
makes the simple approach of quadratically subtracting the HWHM 
of the PSF from the observed one to obtain a measure of the 
intrinsic cluster size unreliable.
Nevertheless, it is already clear from this simple analysis 
that the clusters are very compact, with typical intrinsic
HWHM values of around $5\pc$ (Figure ~\ref{fig:size_hist}b).

As a whole, the \nc s appear to be a very homogeneous class, not only
in luminosity, but also in their structural parameters.
The absence of unresolved nuclear sources
in late-type galaxies - as suggested by this preliminary analysis - 
suggests that any accretion-powered emission from
active galactic nuclei (AGN) is optically weak
in most galaxies of late Hubble types.

\section{Discussion}\label{sec:disc}
\subsection{Are the clusters truly nuclear?}\label{subsec:nuc?}
The question of whether the clusters are indeed located at the 
photometric center of their respective host galaxy is not easily answered,
because the term ``photometric center'' is not well defined itself.
For our analysis, we have defined the photometric center as
the average isophote center of our {\tt ellipse} fit results 
for radii between $2\as$ (well beyond the extent of the cluster)
and that of the outermost fitted ellipse -
in most cases around $15\as$. For the median
distance of our sample, this radial range corresponds to linear
scales between $200\pc$ and $1.5\kpc$. If present, a stellar bar is likely
to dominate the luminosity within this range, but since in
the absence of close interactions a bar
should be symmetric with respect to the dynamical center, its 
photocenter is likely a good measure of the true galaxy nucleus.

Figure~\ref{fig:offsets}a shows a histogram of the projected angular
distance between the position of the presumed nuclear cluster and the
galaxy photocenter according to the definition above. About 75\%
(45 out of 59) clusters lies within $1\as$ from the photocenter. This
angular separation corresponds to about $90\pc$ at the median distance 
of our sample ($19\mpc$). In this representation, some clusters appear
to be well-separated from the galaxy center. However, we caution
that in many galaxies of our sample, the photocenter is poorly
defined, and its position very uncertain.

In order to visualize the uncertainties in the photocenter positions, 
we show in Figure~\ref{fig:offsets}b a plot of the projected offset 
between photocenter and nuclear cluster for all galaxies in group 1. 
Here, the size of the crosses indicates the standard deviation $\sigma$
in the image x and y directions of the isophote centers measured 
from isophotes at radii $\geq 2\arcsec$.
If one assumes that the isophote centers are subject 
to random measurement errors, then the error in the photocenter
position should be $\sigma / \sqrt{n}$, where $n$ is the number of isophotes. 
However, the 
isophotal fits are clearly influenced by dust lanes, extended star 
formation, or other asymmetries that vary with isophote radius, and thus 
make the determination of the photocenter of an individual galaxy 
subject to systematic uncertainties. We therefore have conservatively
estimated the error in the photocenter position to equal $\sigma$, without
dividing by $\sqrt{n}$.

The fact that this estimate is indeed a conservative one is
demonstrated in Figure~\ref{fig:offsets}c which
shows the cumulative distribution of cluster positions 
inside a certain number of standard deviations. The observed sample 
is compared to the expected curve for a two-dimensional normal (i.e. Gaussian)
error distribution. The observed distribution is narrower than
the prediction which indicates that we have somewhat overestimated
our errors in determining the photocenter.

While this analysis does not prove that 
each individual cluster does indeed occupy the true nucleus of its 
host galaxy, the results demonstrate that the
majority of clusters are - within the errors - located at or very near
the photometric center. In the absence of any kinematical information,
it is reasonable to assume that the photocenter coincides
with the dynamical center. We therefore conclude that these clusters 
can rightfully be called ``nuclear''. 

\subsection{Are nuclear clusters special?}
One of the most interesting results from {\it HST} imaging has been the
discovery of extremely luminous, compact, young star clusters in a variety of
starburst environments, including merging galaxies \citep{con94,whi99}, 
dwarf galaxies \citep{hun94,cal97}, and in the circumnuclear rings of 
nearby spiral galaxies \citep{bar95,but00,mao01}. Prior to {\it HST}, only a few 
objects of this type were known to exist \citep{arp85,mel85}; 
the severe crowding in most starbursts made it impossible to resolve 
the individual clusters in ground-based images. Such ``super star clusters''
apparently form preferentially during extreme episodes of violent star 
formation and may be the basic building blocks of starbursts.  Barely 
resolved by {\it HST}, they have effective radii of only $2-4\pc$ and 
luminosities that range as high as $M_V = -14$ to $-15$ mag.  The small 
radii, high luminosities, and presumably high masses of these clusters have 
led to suggestions that they may remain as bound systems and therefore could 
be present-day versions of young globular clusters \citep[e.g.][]{ho96}.  

The nuclear clusters discovered in our survey bear a close resemblance to 
off-nuclear super star clusters.  Although we do not yet have definitive 
size measurements for our sources, the observed HWHM values range from 
$\sim$1 to 10 pc, with a median value of $\sim$5 pc; this is consistent with
the sizes of super star clusters.  Similarly, the optical absolute 
magnitudes of the nuclear clusters lie comfortably within the luminosity 
function of super star clusters.
\section{Conclusions}\label{sec:concl}
We have presented a catalog of {\it HST}/WFPC2 I-band images of an
unbiased sample of 77 nearby, late-type spiral galaxies with
low inclination. From an isophotal analysis of
the images, we demonstrate that about 75\% of the sample galaxies
host a compact, luminous stellar cluster at or very close to their
photocenter. These clusters often are completely isolated from
other comparable structures, emphasizing that even in the 
relatively shallow potential wells of late-type galaxy disks, the
center is well-defined, and has a unique star formation history.
From analytical fits to the surface brightness profiles,
we determine the flux attributable to the cluster. The distribution
of absolute cluster luminosities has a FWHM of 4 magnitudes, and 
a median value of $\rm M_I = -11.5$, comparable to young super star clusters
in starbursting galaxies. Together with initial estimates of their
size distribution, this suggests that \nc s in spiral galaxies of the latest
Hubble types are a fairly homogenous class of objects.
The dataset is a representative survey of late-type spiral galaxies in
the local universe, and as such yields a valuable source catalog for
spectroscopic follow-up observations which are needed to further
constrain the star formation history of \nc s. We have begun 
such a follow-up program both with {\it HST} and ground-based
observatories.

\acknowledgements
The anonymous referee provided useful comments
which helped to improve the presentation of this paper.
Support for proposal \#\,8599 was provided by NASA through a grant from 
the Space Telescope Science Institute, which is operated by the Association 
for Research in Astronomy, Inc., under NASA contract no. NAS 5-26555.
This research has made use of the NASA/IPAC Extragalactic Database (NED) 
which is operated by the Jet Propulsion Laboratory, California Institute of 
Technology, under contract with NASA. 
It has also benefited greatly from use of the Lyon-Meudon
Extragalactic Database (LEDA, http://leda.univ-lyon1.fr).


\newpage
\figcaption[boeker.f1.eps]{\label{fig:maps1}
   WFPC2 F814W images of the 59 galaxies with evidence for a nuclear
   cluster. The bar in the upper
   left corner represents a scale of $1\kpc$, calculated from the
   distances listed in Table~\ref{tbl:res}. For a few very nearby
   objects, the bar is dashed, in which case it indicates
   a scale of $250\pc$. The symbol in the top right corner
   indicates north (with arrow) and east directions. 
   All images are on a logarithmic grey-scale stretch,
   optimized for the dynamic range of the galaxy. The object identified 
   as the central cluster is circled in sources where a visual 
   identification may be ambiguous.
   }
\figcaption[boeker.f2.eps]{\label{fig:maps2}
   As Figure~\ref{fig:maps1} for the 18 galaxies without evidence for a nuclear
   cluster. 
   }
\figcaption[boeker.f3.eps]{\label{fig:profiles}
   I-band surface brightness profiles (SBP) of the 59 galaxies 
   with evidence for a nuclear cluster. The 
   diamond-shaped symbols indicate the results of the elliptical isophote
   fits. The formal error bars of the isophote fits are also shown; in
   most cases, they are contained within the symbol size.
   The lines show the best fit analytical model to the inner part 
   of the SBP as described in \S~\ref{sec:obs} (dotted), the 
   inward-extrapolated best fit outside of the \nc\ (solid), and the 
   constant surface brightness level (dashed) at the radius where 
   the SBP starts to deviate from the pure disk profile. 
   This radius was used to derive the cluster luminosity as 
   described in \S~\ref{subsec:phot}.
   }
\figcaption[boeker.f4.eps]{\label{fig:badprofiles}
   As Figure~\ref{fig:profiles} for those 14 galaxies in group 2
   (without evidence for a \nc ) for which the isophotal fit was
   successful.
   }
\figcaption[boeker.f5.eps]{\label{fig:hist}
   Histogram of apparent (left) and absolute (right)
   I-band magnitudes of all identified \nc s.
   Also shown are the median absolute luminosity of the
   sample (asterisk) and the luminosity of a cluster
   with $10^6$ and $10^7\lsun$, respectively (diamonds).
   }
\figcaption[boeker.f6.eps]{\label{fig:size_hist}
   Histogram of angular (left) and linear (right)
   observed HWHM radius (i.e. not corrected for PSF convolution) 
   of all identified \nc s. The vertical dashed line in the
   left panel denotes the HWHM radius of the F814W PSF.
   }
\figcaption[boeker.f7.eps]{\label{fig:offsets}
   a) distribution of projected distances between \nc\ position and the
   average isophote center between $2\as$ and $15\as$. b) projected
   position of all \nc s, relative to the photocenter
   of their respective host galaxy. The size of the crosses
   denotes the 1$\sigma$ scatter in the isophote centers.
   c) cumulative distribution of projected distance between cluster
   and photocenter (solid curve), compared to the expected curve
   for a two-dimensional Gaussian error distribution (dotted curve). 
   }
\newpage
\begin{deluxetable}{lcccccccc}
\tabletypesize{\scriptsize}
\tablecaption{Summary of Observations \label{tbl:obs}}
\tablewidth{0pt}
\tablehead{
\colhead{(1)} & \colhead{(2)} & \colhead{(3)} & \colhead{(4)} &
\colhead{(5)} & \colhead{(6)} & \colhead{(7)} & \colhead{(8)} & \colhead{(9)} \\
\colhead{Galaxy} & \colhead{R.A.}  & \colhead{Dec.} & \colhead{$\rm v_z$} & 
\colhead{Type} & \colhead{$\rm m_B$} & \colhead{$\rm A_I$} & 
\colhead{$\rm d_{MA}$} & \colhead{Obs. date} \\
 & \colhead{(J2000)} & \colhead{(J2000)} & \colhead{[km/s]} &  & \colhead{[mag]} & 
 \colhead{[mag]} & \colhead{[arcmin]} &   
}
\startdata
NGC\,275    & 00 51 04.20 & -07 04 00.0 & 1681 & SB(rs)cd pec & 13.16 & 0.109 & 1.5 & 07/09/01  \\
NGC\,300    & 00 54 53.47 & -37 41 00.0 &  -54 & SA(s)d    & 8.95  & 0.025 & 21.9 & 05/06/01 \\
NGC\,337a   & 01 01 33.90 & -07 35 17.7 &  998 & SAB(s)dm & 14.92 & 0.189 & 5.9 & 07/10/01  \\
NGC\,428    & 01 12 55.60 & -00 58 54.4 & 1130 & SAB(s)m   & 11.91 & 0.055 & 4.1 & 01/06/01  \\
NGC\,450    & 01 15 30.52 & -00 51 38.3 & 1720 & SAB(s)cd: & 12.20 & 0.077 & 3.1 & 07/12/01  \\
ESO\,80-6   & 01 47 16.87 & -62 58 14.8 & 1227 & SB(s)m    & 14.37 & 0.052 & 1.4 & 07/10/01  \\
NGC\,600    & 01 33 05.25 & -07 18 42.1 & 1763 & SB(rs)d   & 13.65 & 0.073 & 3.3 & 07/09/01  \\
NGC\,853    & 02 11 43.35 & -09 18 01.1 & 1413 & Sm pec    & 13.43 & 0.050 & 1.5 & 07/10/00  \\
NGC\,1042   & 02 40 23.63 & -08 25 59.8 & 1271 & SAB(rs)cd & 12.50 & 0.056 & 4.7 & 01/26/01  \\
NGC\,1313   & 03 18 15.37 & -66 29 50.6 &  174 & SB(s)d    &  9.20 & 0.212 & 9.1 & 01/12/01  \\
ESO\,358-5  & 03 27 16.47 & -33 29 06.1 & 1409 & SAB(s)m pec: & 14.90 & 0.022 & 1.4 & 05/31/01  \\
ESO\,418-8  & 03 31 30.48 & -30 12 44.6 &  988 & SB(r)d    & 13.68 & 0.029 & 1.2 & 05/30/01  \\
NGC\,1493   & 03 57 27.73 & -46 12 38.1 &  796 & SB(rs)cd  & 11.78 & 0.020 & 3.5 & 05/02/01  \\
ESO\,202-41 & 04 36 56.69 & -52 10 25.2 & 1396 & SB(s)m    & 14.94 & 0.017 & 1.2 & 04/27/01  \\
ESO\,85-47  & 05 07 43.86 & -62 59 24.3 & 1180 & SB(s)m    & 14.53 & 0.050 & 1.7 & 04/26/01  \\
ESO\,204-22 & 05 36 26.06 & -52 11 02.5 & 1005 & SB(s)m: pec & 15.44 & 0.080 & 1.3 & 12/25/00 \\
NGC\,2139   & 06 01 07.90 & -23 40 21.3 & 1649 & SAB(rs)cd & 11.99 & 0.065 & 2.6 & 02/15/01  \\
UGC\,3574   & 06 53 10.60 & +57 10 39.0 & 1635 & SA(s)cd   & 13.20 & 0.103 & 4.2 & 01/30/01  \\  
UGC\,3826   & 07 24 32.05 & +61 41 35.2 & 1946 & SAB(s)d   & 14.10 & 0.133 & 3.5 & 10/06/00   \\
NGC\,2552   & 08 19 20.14 & +50 00 25.2 &  695 & SA(s)m?   & 12.56 & 0.090 & 3.5 & 04/06/01  \\
UGC\,4499   & 08 37 41.43 & +51 39 11.1 &  877 & SAdm      & 13.50 & 0.069 & 2.6 & 05/31/01  \\
NGC\,2763   & 09 06 49.26 & -15 29 59.9 & 1769 & SB(r)cd pec & 12.64 & 0.141 & 2.3 & 04/02/01  \\
NGC\,2805   & 09 20 24.56 & +64 05 55.2 & 1968 & SAB(rs)d  & 11.52 & 0.100 & 6.3 & 10/07/00   \\
UGC\,4988   & 09 23 15.26 & +34 44 03.7 & 1696 & SABm      & 15.30 & 0.036 & 1.1 & 06/03/01  \\
UGC\,5015   & 09 25 47.89 & +34 16 35.9 & 1800 & SABdm     & 14.90 & 0.034 & 1.9 & 06/04/01  \\
UGC\,5288   & 09 51 17.00 & +07 49 39.0 &  559 & Sdm:      & 14.09 & 0.066 & 1.3 & 01/14/01  \\
NGC\,3206   & 10 21 47.65 & +56 55 49.6 & 1380 & SB(s)cd   & 12.57 & 0.027 & 3.0 & 05/14/01  \\
NGC\,3346   & 10 43 38.90 & +14 52 18.0 & 1315 & SB(rs)cd  & 12.41 & 0.054 & 2.9 & 01/14/01  \\
NGC\,3423   & 10 51 14.30 & +05 50 24.0 & 1025 & SA(s)cd   & 11.59 & 0.058 & 3.8 & 02/07/01  \\
NGC\,3445   & 10 54 35.87 & +56 59 24.4 & 2245 & SAB(s)m   & 12.90 & 0.015 & 1.6 & 06/01/01   \\
NGC\,3782   & 11 39 20.72 & +46 30 48.6 &  944 & SAB(s)cd: & 13.10 & 0.035 & 1.7 & 05/10/01  \\
NGC\,3906   & 11 49 40.46 & +48 25 33.3 & 1166 & SB(s)d    & 13.49 & 0.050 & 1.9 & 03/09/01  \\
NGC\,3913   & 11 50 38.77 & +55 21 12.1 & 1190 & SA(rs)d:  & 13.17 & 0.025 & 2.6 & 01/17/01  \\
A\,1156+52  & 11 59 09.47 & +52 42 26.1	& 1307 & SB(rs)cd  & 13.12 & 0.053 & 3.5 & 08/03/01  \\
ESO\,504-30 & 11 57 15.14 & -27 42 00.2 & 1673 & SB(r)d:   & 14.66 & 0.142 & 1.1 & 05/06/01  \\
UGC\,6931   & 11 57 22.79 & +57 55 22.5 & 1446 & SBm:      & 14.31 & 0.049 & 1.4 & 08/16/00   \\
NGC\,4027   & 11 59 30.50 & -19 15 44.0 & 1588 & SB(s)dm   & 11.66 & 0.081 & 3.2 & 04/07/01   \\
NGC\,4204   & 12 15 14.51 & +20 39 30.7 &  968 & SB(s)dm   & 12.90 & 0.065 & 3.6 & 06/11/01   \\ 
NGC\,4299   & 12 21 40.90 & +11 30 03.0 &  306 & SAB(s)dm: & 12.88 & 0.063 & 1.7 & 04/28/01   \\ 
NGC\,4416   & 12 26 46.72 & +07 55 07.9 & 1449 & SB(rs)cd: & 13.14 & 0.049 & 1.7 & 03/19/01   \\
NGC\,4411B  & 12 26 47.30 & +08 53 04.5 & 1334 & SAB(s)cd  & 12.91 & 0.058 & 2.5 & 05/28/01   \\
NGC\,4487   & 12 31 04.36 & -08 03 13.8 & 1020 & SAB(rs)cd & 12.26 & 0.041 & 4.2 & 05/28/01   \\
NGC\,4496A  & 12 31 39.32 & +03 56 22.7 & 1772 & SB(rs)m   & 11.94 & 0.048 & 4.0 & 03/17/01   \\
NGC\,4517A  & 12 32 28.15 & +00 23 22.8 & 1554 & SB(rs)dm: & 12.94 & 0.046 & 4.0 & 03/18/01  \\
NGC\,4540   & 12 34 50.90 & +15 33 06.9 & 1383 & SAB(rs)cd & 12.44 & 0.065 & 1.9 & 07/19/00   \\
NGC\,4618   & 12 41 32.74 & +41 09 03.8 &  748 & SB(rs)m   & 11.22 & 0.041 & 4.2 & 07/10/01  \\
NGC\,4625   & 12 41 52.61 & +41 16 26.3 &  816 & SAB(rs)m pec & 12.92 & 0.035 & 2.2 & 05/28/01   \\
NGC\,4701   & 12 49 11.71 & +03 23 21.8 &  768 & SA(s)cd   & 12.80 & 0.057 & 2.8 & 05/29/01   \\
NGC\,4775   & 12 53 45.79 & -06 37 20.1 & 1565 & SA(s)d    & 12.24 & 0.067 & 2.1 & 12/21/00   \\
NGC\,4904   & 13 00 56.97 & -00 01 31.9 & 1204 & SB(s)cd   & 12.60 & 0.050 & 2.2 & 07/12/00   \\
A\,1301-03  & 13 04 31.43 & -03 34 20.3 & 1379 & SAB(s)dm  & 12.90 & 0.058 & 3.5 & 06/02/01   \\
IC\,4182    & 13 05 49.53 & +37 36 17.6 &  515 & SA(s)m    & 13.0  & 0.027 & 6.0 & 07/09/01  \\
ESO\,444-2  & 13 16 44.91 & -27 53 09.7 & 1544 & SAB(s)dm  & 14.97 & 0.142 & 1.1 & 05/30/01   \\
NGC\,5068   & 13 18 54.60 & -21 02 19.7 &  607 & SB(s)d    & 10.52 & 0.197 & 7.2 & 06/02/01   \\
UGC\,8516   & 13 31 52.50 & +20 00 01.0 & 1156 & Scd:      & 14.03 & 0.057 & 1.1 & 06/01/01   \\
ESO\,510-59 & 14 04 46.43 & -24 49 40.7 & 2267 & SB(s)cd   & 13.61 & 0.138 & 2.5 & 04/27/01   \\
NGC\,5477   & 14 05 31.25 & +54 27 12.3 &  565 & SA(s)m    & 14.36 & 0.021 & 1.7 & 09/22/00   \\
NGC\,5585   & 14 19 48.08 & +56 43 43.8 &  571 & SAB(s)d   & 11.20 & 0.030 & 5.8 & 05/05/01  \\
NGC\,5584   & 14 22 23.65 & -00 23 09.2 & 1695 & SAB(rs)cd & 12.63 & 0.075 & 3.4 & 04/18/01  \\
NGC\,5668   & 14 33 24.30 & +04 27 02.0 & 1665 & SA(s)d    & 12.2  & 0.071 & 3.3 & 04/23/01  \\
NGC\,5669   & 14 32 44.00 & +09 53 31.0	& 1481 & SAB(rs)cd & 12.03 & 0.053 & 4.0 & 07/14/01  \\
NGC\,5774   & 14 53 42.60 & +03 34 59.0 & 1648 & SAB(rs)d  & 12.74 & 0.081 & 3.0 & 05/20/01  \\
NGC\,5789   & 14 56 35.52 & +30 14 02.5 & 2002 & Sdm	   & 14.70 & 0.041 & 0.9 & 03/17/01  \\
NGC\,5964   & 15 37 36.30 & +05 58 26.0 & 1552 & SB(rs)d   & 12.6  & 0.113 & 4.2 & 05/02/01  \\
ESO\,138-10 & 16 59 02.96 & -60 12 02.9 &  942 & SA(s)cd   & 11.59 & 0.427 & 5.6 & 05/29/01   \\
NGC\,6509   & 17 59 25.36 & +06 17 12.4 & 1926 & Sd        & 13.10 & 0.375 & 1.6 & 07/03/00   \\
NGC\,6946   & 20 34 52.34 & +60 09 14.2 &  310 & SAB(rs)cd &  9.61 & 0.663 & 11.5 & 12/03/00   \\
ESO\,187-51 & 21 07 33.09 & -54 57 02.0 & 1158 & SB(s)m	   & 14.85 & 0.066 & 1.9 & 03/23/01   \\
UGC\,12082  & 22 34 11.54 & +32 52 10.3 &  974 & Sm        & 14.1  & 0.185 & 2.6 & 08/13/00   \\
NGC\,7418   & 22 56 36.00 & -37 01 47.3 & 1287 & SAB(rs)cd & 12.30 & 0.031 & 3.5 & 06/01/01   \\
NGC\,7424   & 22 57 18.08 & -41 04 19.0 &  765 & SAB(rs)cd & 10.96 & 0.021 & 9.5 & 06/01/01   \\
ESO\,290-39 & 23 03 29.14 & -46 02 22.8 & 1337 & SB(s)m    & 15.0  & 0.028 & 1.1 & 10/12/00   \\
UGC\,12732  & 23 40.39.80 & +26 14 10.0 &  870 & Sm:       & 13.8  & 0.172 & 3.0 & 05/14/01   \\
ESO\,241-6  & 23 56 13.08 & -43 26 00.0 & 1219 & SB(s)m    & 14.4  & 0.025 & 1.1 & 11/12/00   \\
NGC\,7689   & 23 33 16.11 & -54 05 37.0	& 1744 & SA(r)c    & 12.2  & 0.023 & 2.9 & 07/12/01   \\
NGC\,7741   & 23 43 53.65 & +26 04 33.1 &  872 & SB(s)cd   & 11.84 & 0.145 & 4.4 & 07/24/01   \\
NGC\,7793   & 23 57 49.75 & -32 35 29.5 &   69 & SA(s)d    &  9.98 & 0.038 & 9.3 & 04/19/01
\enddata
\tablecomments{Columns 1-3: object name and coordinates, as taken from
the NASA Extragalactic Database (NED). Column 4: recession velocity,
corrected according to the Virgo-centric infall model \citep{san90},
taken from the Lyon-Meudon Extragalactic Database (LEDA).
Columns 5 and 6: galaxy morphological type and apparent total B-magnitude
(NED). Column 7: Galactic foreground extinction \citep{sch98}, converted
to I-band using the \cite{car89} extinction law, and $\rm R_V = 3.1$ (NED).
Column 8: galaxy major axis diameter (NED). Column 9: date of observation.
}
\end{deluxetable}
\newpage
\begin{deluxetable}{lcccccccc}
\tabletypesize{\scriptsize}
\tablecaption{Nuclear Cluster Properties \label{tbl:res}}
\tablewidth{0pt}
\tablehead{
\colhead{(1)} & \colhead{(2)} & \colhead{(3)} & \colhead{(4)} &
\colhead{(5)} & \colhead{(6)} & \colhead{(7)} & \colhead{(8)} & \colhead{(9)}\\
\colhead{Galaxy} & \colhead{Distance} & \colhead{$\rm HWHM$} & \colhead{HWHM} & \colhead{$R_u$} &
\colhead{$\rm m_I$} & \colhead{$\rm M_I$} & \colhead{$\mu _0$} & \colhead{Type of fit} \\
 & \colhead{[Mpc]} & \colhead{[$\as$]} & \colhead{[pc]} & \colhead{[$\as$]} & \colhead{[mag]} & 
 \colhead{[mag]} & \colhead{[mag]} & 
}
\startdata
NGC\,275    & 24.0  &   0.054  &     6.3  & 0.5 & 19.47$\pm$0.01 & -12.54 & 15.312 & u  \\
NGC\,300    &  2.2$^1$ & 0.133 &     1.4  & 2.5 & 15.29$\pm$0.40 & - 11.43 & 13.651 & u  \\
NGC\,337a   & 14.3  &   0.058  &     4.0  & 0.6 & 20.94$\pm$0.01 & -10.02 & 17.144 & u$^*$  \\
NGC\,428    & 16.1  &   0.046  &     3.6  & 0.9 & 17.95$\pm$0.01 & -13.15 & 13.875 & u  \\
NGC\,450    & 25.6  &	0.112  &    13.3  & 0.4 & 20.13$\pm$0.17 & -11.90 & 16.584 & u  \\
ESO\,80-6   & 17.5  & - & - & - & - & - & - & fc  \\ 
NGC\,600    & 25.2  &   0.057  &     7.0  & 0.3 & 19.92$\pm$0.03 & -12.16 & 15.712 & u  \\
NGC\,853    & 20.2  &   0.054  &     5.3  & 0.25 & 19.90$\pm$0.04 & -11.68 & 15.515 & u  \\
NGC\,1042   & 18.2  &   0.052  &     4.6  & 0.2 & 18.40$\pm$0.29 & -12.95 & 13.464 & u \\
NGC\,1313   &  4.4$^2$  & - & - & - & - & - & - & fc  \\
ESO\,358-5  & 20.1  &   0.055  &     5.4  & 1.0 & 20.10$\pm$0.06 & -11.44 & 16.358 & u$^*$  \\ 
ESO\,418-8  & 14.1  &   0.052  &     3.6  & 0.3 & 20.54$\pm$0.01 & -10.24 & 16.154 & u  \\ 
NGC\,1493   & 11.4  &   0.058  &     3.2  & 0.5 & 17.17$\pm$0.03 & -13.13 & 13.259 & u  \\
ESO\,202-41 & 19.9  &   0.059  &     5.7  & 0.4 & 22.51$\pm$0.03 & -9.01 & 18.277 & u$^*$  \\ 
ESO\,85-47  & 16.9  & - & - &-  & - & - & - & fc  \\ 
ESO\,204-22 & 14.4  &- & - & - & - & - & - & fc  \\ 
NGC\,2139   & 23.6  &   0.066  &     7.5  & 0.25 & 19.28$\pm$0.29 & -12.65 & 14.652 & u  \\
UGC\,3574   & 23.4  &   0.065  &     7.4  & 0.5 & 20.04$\pm$0.18 & -11.90 & 16.063 & u  \\  
UGC\,3826   & 27.8  &   0.074  &    10.0  & 0.2 & 21.60$\pm$0.02 & -10.76 & 17.375 & u \\  
NGC\,2552   &  9.9  &   0.053  &     2.6  & 1.0 & 18.04$\pm$0.01 & -12.04 & 14.220 & u  \\ 
UGC\,4499   & 12.5  &   0.072  &     4.4  & 0.3 & 21.97$\pm$0.63 & -8.59 & 17.902 & u$^*$  \\  
NGC\,2763   & 25.3  &   0.067  &     8.2  & 0.2 & 20.59$\pm$0.35 & -11.56 & 15.350 & u  \\ 
NGC\,2805   & 28.1  &   0.060  &     8.2  & 0.5 & 19.02$\pm$0.06 & -13.32 & 14.987 & u  \\  
UGC\,4988   & 24.2  &   0.054  &     6.4  & 0.3 & 20.76$\pm$0.04 & -11.20 & 16.397 & u  \\  
UGC\,5015   & 25.7  &   0.065  &     8.2  & 0.3 & 20.71$\pm$0.01 & -11.37 & 16.686 & u  \\  
UGC\,5288   &  8.0  & - & - & - & - & - & - & fc  \\  
NGC\,3206   & 19.7  & - & - & - & - & - & - & fc \\ 
NGC\,3346   & 18.8  &   0.042  &     3.8  & 0.3 & 19.64$\pm$0.01 & -11.78 & 15.106 & u  \\ 
NGC\,3423   & 14.6  &   0.057  &     4.1  & 0.3 & 19.04$\pm$0.05 & -11.84 & 14.876 & u  \\ 
NGC\,3445   & 32.1  &   0.051  &     7.9  & 0.4 & 19.12$\pm$0.10 & -13.42 & 14.794 & u  \\  
NGC\,3782   & 13.5  &   0.055  &     3.6  & 0.25 & 20.61$\pm$0.01 & -10.07 & 16.134 & u  \\  
NGC\,3906   & 16.7  &   0.059  &     4.8  & 0.3 & 21.15$\pm$0.11 & -10.01 & 16.759 & u  \\ 
NGC\,3913   & 17.0  &   0.255  &    21.0  & 0.3 & 21.22$\pm$0.07 & -9.96 & 17.256 & u  \\ 
A\,1156+52  & 18.7  &	0.055  &     5.0  & 0.4 & 20.43$\pm$0.01 & -10.98 & 16.19 & u	\\
ESO\,504-30 & 23.9  &   0.056  &     6.5  & 0.3 & 20.70$\pm$0.05 & -11.33 & 16.522 & u  \\  
UGC\,6931   & 20.7  &   0.052  &     5.2  & 0.5 & 21.91$\pm$0.12 & -9.72 & 17.757 & u$^\prime$  \\  
NGC\,4027   & 22.7  &   0.066  &     7.3  & 0.2 & 20.38$\pm$0.22 & -11.48 & 15.335 & u  \\  
NGC\,4204   & 13.8  &   0.066  &     4.5  & 0.4 & 20.51$\pm$0.02 & -10.26 & 16.590 & u$^*$  \\  
NGC\,4299   & 16.8$^3$ & 0.051 &     1.1  & 0.25 & 19.46$\pm$0.04 & -11.73 & 14.912 & u  \\  
NGC\,4416   & 20.7  &   0.219  &    22.0  & 0.25 & 22.82$\pm$0.64 & -8.81 & 17.415 & u$^\prime$  \\  
NGC\,4411B  & 19.1  &   0.062  &     5.8  & 0.3 & 18.89$\pm$0.07 & -12.57 & 14.892 & u  \\  
NGC\,4487   & 14.6  &   0.051  &     3.6  & 0.25 & 17.89$\pm$0.01 & -12.97 & 13.391 & u  \\  
NGC\,4496A  & 25.3  &   0.048  &     6.0  & 0.3 & 20.08$\pm$0.02 & -11.99 & 15.631 & u$^\prime$  \\  
NGC\,4517A  & 22.2  & - & - & - & - & - & - & fc  \\  
NGC\,4540   & 19.8  &   0.069  &     6.6  & 0.3 & 19.25$\pm$0.02 & -12.29 & 15.098 & u  \\  
NGC\,4618   & 10.7  &   0.098  &     5.1  & 0.5 & 18.74$\pm$0.06 & -11.45 & 15.668 & u \\
NGC\,4625   & 11.7  &   0.097  &     5.5  & 0.3 & 19.76$\pm$0.08 & -10.61 & 15.803 & u  \\
NGC\,4701   & 11.0  &   0.044  &     2.4  & 0.5 & 16.81$\pm$0.07 & -13.45 & 12.641 & u  \\
NGC\,4775   & 22.4  &   0.056  &     6.1  & 0.3 & 18.04$\pm$0.05 & -13.77 & 13.852 & u  \\  
NGC\,4904   & 17.2  & - & - & - & - & - & - & fc  \\  
A\,1301-03  & 19.7  & - & - & - & - & - & - & ngf  \\  
IC\,4182    &  7.4  & - & - & - & - & - & - & ngf \\
ESO\,444-2  & 22.1  & - & - & - & - & - & - & fc  \\  
NGC\,5068   &  8.7  &   0.106  &     4.5  & 1.0 & 17.55$\pm$0.05 & -12.34 & 15.194 &  u  \\  
UGC\,8516   & 16.5  &   0.089  &     7.2  & 0.4 & 20.18$\pm$0.09 & -10.97 & 16.615 & u  \\ 
ESO\,510-59 & 32.4  & - & - & - & - & - & - & ngf  \\  
NGC\,5477   &  8.1  & - & - & - & - & - & - & fc  \\  
NGC\,5585   &  8.2  &   0.063  &     2.5  & 0.5 & 18.24$\pm$0.03 & -11.35 & 14.531 & u  \\  
NGC\,5584   & 24.2  &   0.075  &     8.8  & 0.2 & 22.53$\pm$0.58 & -9.47 & 16.837 & u  \\  
NGC\,5789   & 28.6  & - & - & - & - & - & - & fc  \\  
NGC\,5668   & 23.8  &   0.054  &     6.3  & 0.4 & 18.86$\pm$0.06 & -13.10 & 14.757 & u  \\  
NGC\,5669   & 21.2  &	0.123  &    12.6  & 0.25 & 21.66$\pm$0.01 & -10.03 & 17.354 & u$^*$ \\
NGC\,5774   & 23.5  &   0.073  &     8.3  & 0.2 & 21.97$\pm$0.05 & -9.97 & 17.369 & u  \\  
NGC\,5964   & 22.2  &   0.056  &     6.1  & 0.5 & 19.22$\pm$0.06 & -12.62 & 15.210 & u  \\  
ESO\,138-10 & 13.5  &   0.085  &     5.6  & 0.5 & 16.68$\pm$0.13 & -14.40 & 13.367 & u  \\  
NGC\,6509   & 27.5  &   0.043  &     5.8  & 0.25 & 19.49$\pm$0.07 & -13.08 & 14.752 & u  \\  
NGC\,6946   &  5.5$^3$  & - & - & - & - & - &  & ngf  \\  
ESO\,187-51 & 16.5  & - & - & - & - & - & - & fc  \\ 
UGC\,12082  & 13.9  & - & - & - & - & - & - & fc  \\  
ESO\,290-39 & 19.1  &   0.070  &     6.5  & 0.3 & 22.52$\pm$0.01 & -8.92 & 18.261 & u$^\prime$ \\ 
NGC\,7418   & 18.4  &   0.065  &     5.8  & 3.0 & 15.12$\pm$0.22 & -16.23 & 13.027 & u  \\  
NGC\,7424   & 10.9  &   0.097  &     5.1  & 0.5 & 18.80$\pm$0.05 & -11.41 & 15.650 & u  \\  
UGC\,12732  & 12.4  &   0.067  &     4.0  & 1.0 & 19.35$\pm$0.01 & -11.29 & 15.888 & u$^\prime$  \\ 
ESO\,241-6  & 17.4  &   0.056  &     4.7  & 0.3 & 21.30$\pm$0.15 & -9.93 & 16.884 & u  \\
NGC\,7689   & 24.9  &	0.080  &     9.6  & 0.4 & 18.26$\pm$0.12 & -13.75 & 14.576 & u \\
NGC\,7741   & 12.5  & - & - & - & - & - & - & fc \\
NGC\,7793   & 3.3$^4$ & 0.096  &     1.5  & 4.0 & 14.00$\pm$0.03 & -13.64 & 12.551 & u 
\enddata
\tablecomments{Columns 1: object name. Column~2: distance, derived
from the recession velocity in Column~4 of Table~\ref{tbl:obs} and
assuming $\rm H_o = 70\kms$, if not noted otherwise. 
Columns~3 and 4: angular and physical observed
half width at half maximum of the nuclear cluster (i.e. not
corrected for PSF convolution). The conversion
assumes the distances listed in Column~2. Column~5: aperture
radius used to derive cluster luminosity. Column~6: apparent
I-band magnitude of the nuclear cluster. Listed is the average
value for the two background models as described in \S~\ref{subsec:phot},
and half their difference as the uncertainty. Column~7: absolute I-band 
magnitude of the nuclear cluster, corrected for Galactic extinction
as listed in Column~7 of Table~\ref{tbl:obs}. The distance
modulus was derived from the distances in Column~2. Column 8:
peak observed I-band surface brightness in a $0\farcs 0455$ square
pixel (i.e. not corrected for PSF convolution). Column~9: 
type of the isophotal fit: (u) - unconstrained, sometimes with
increased isophote spacing (indicated by $^*$), or over
two separate radial ranges (indicated by $^\prime$) (fc) - fixed ellipse
center and ellipticity, (ngf) - no good fit.
}

\tablerefs{
(1) \cite{fre92}, (2) \cite{dev63}, (3) \cite{tul88}, (4) \cite{car85}
}
\end{deluxetable}

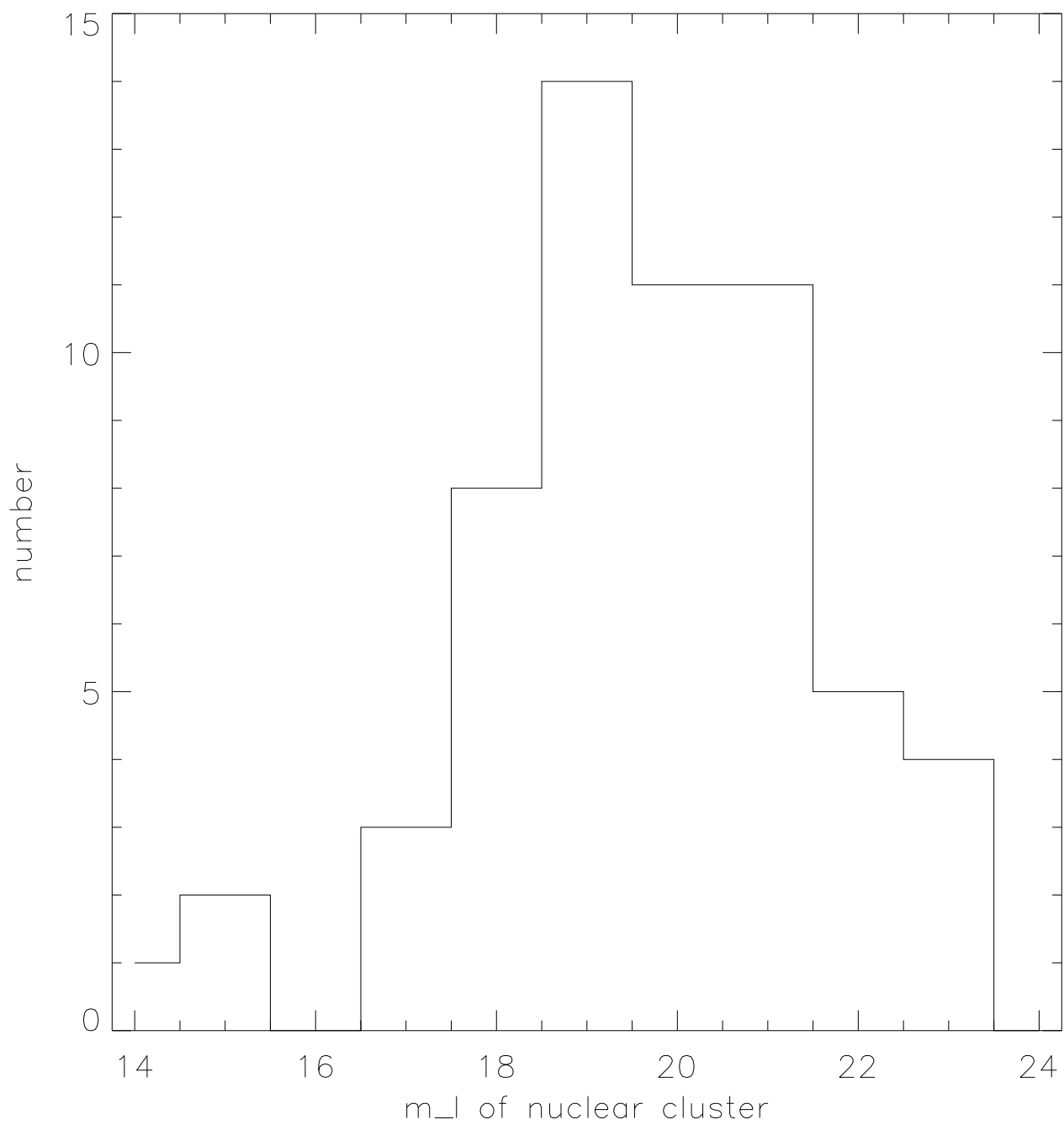
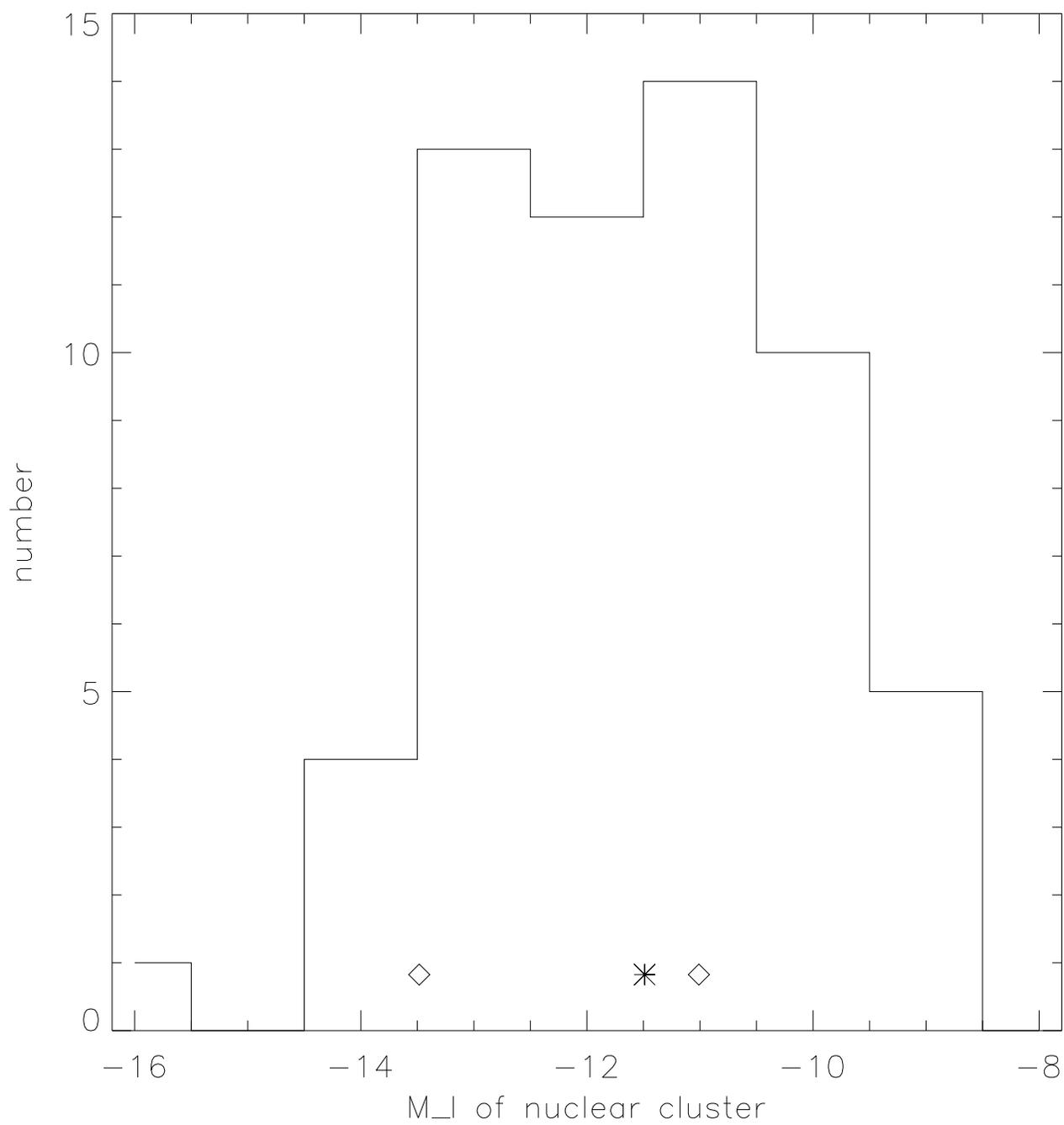
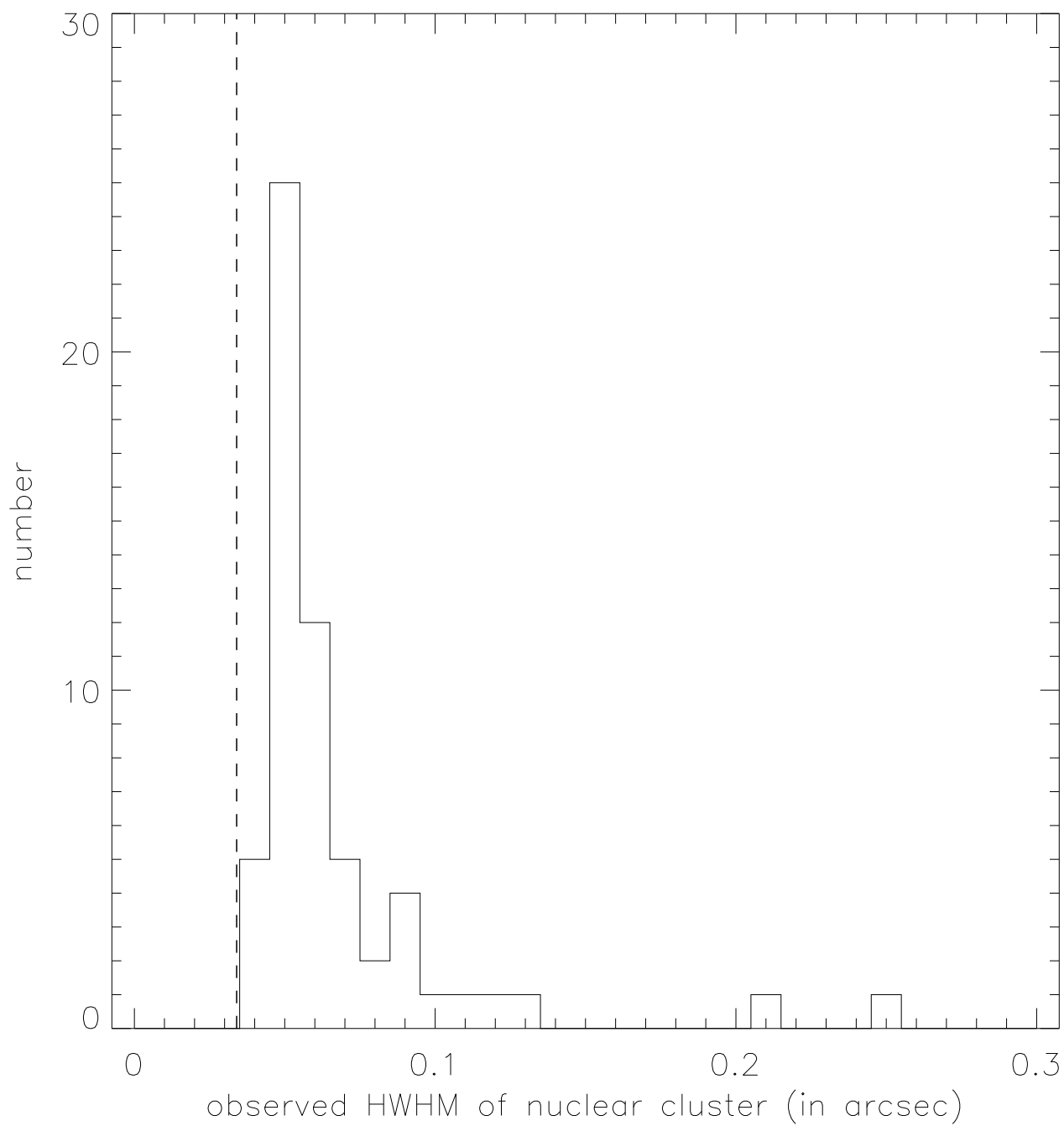
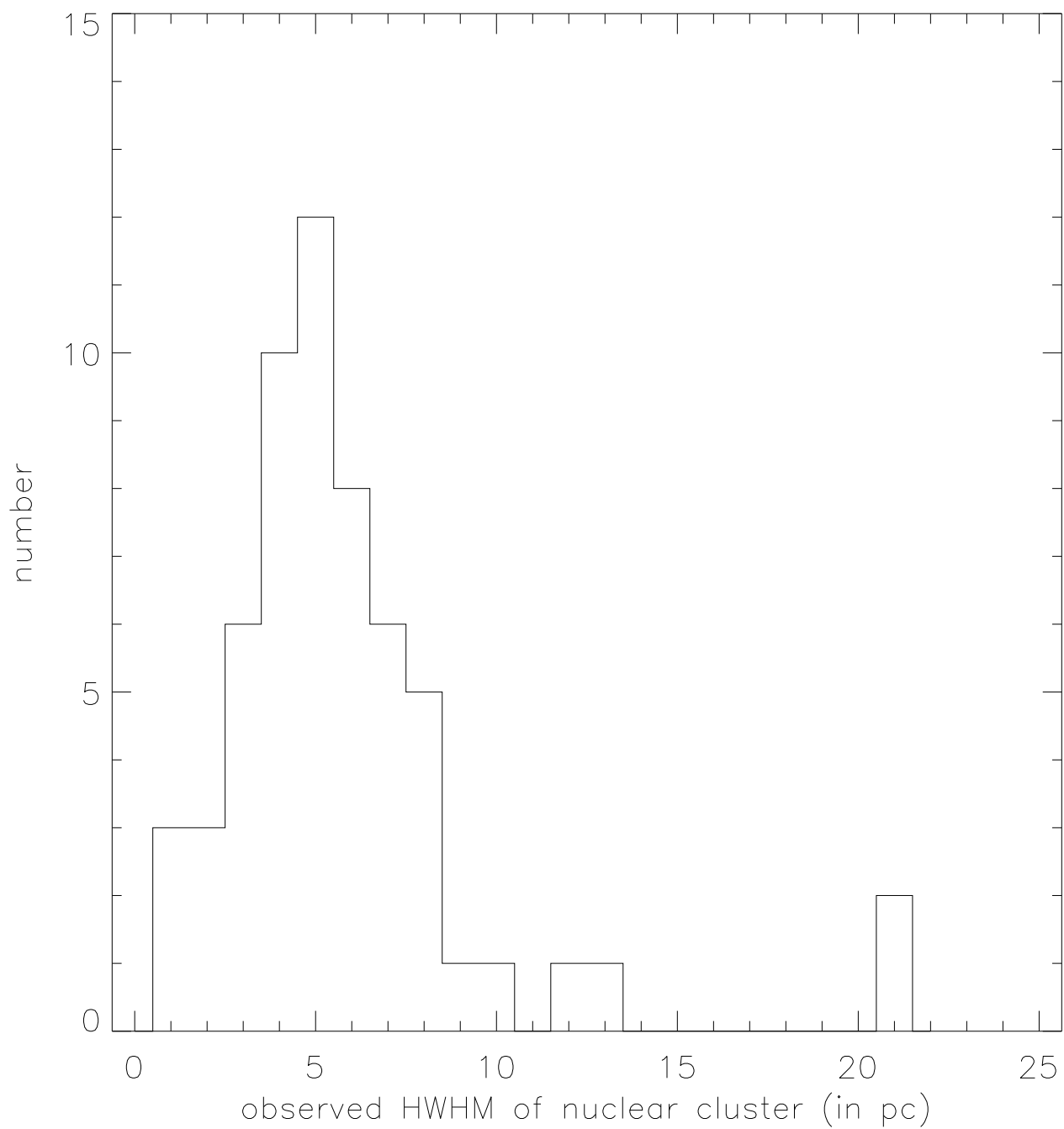
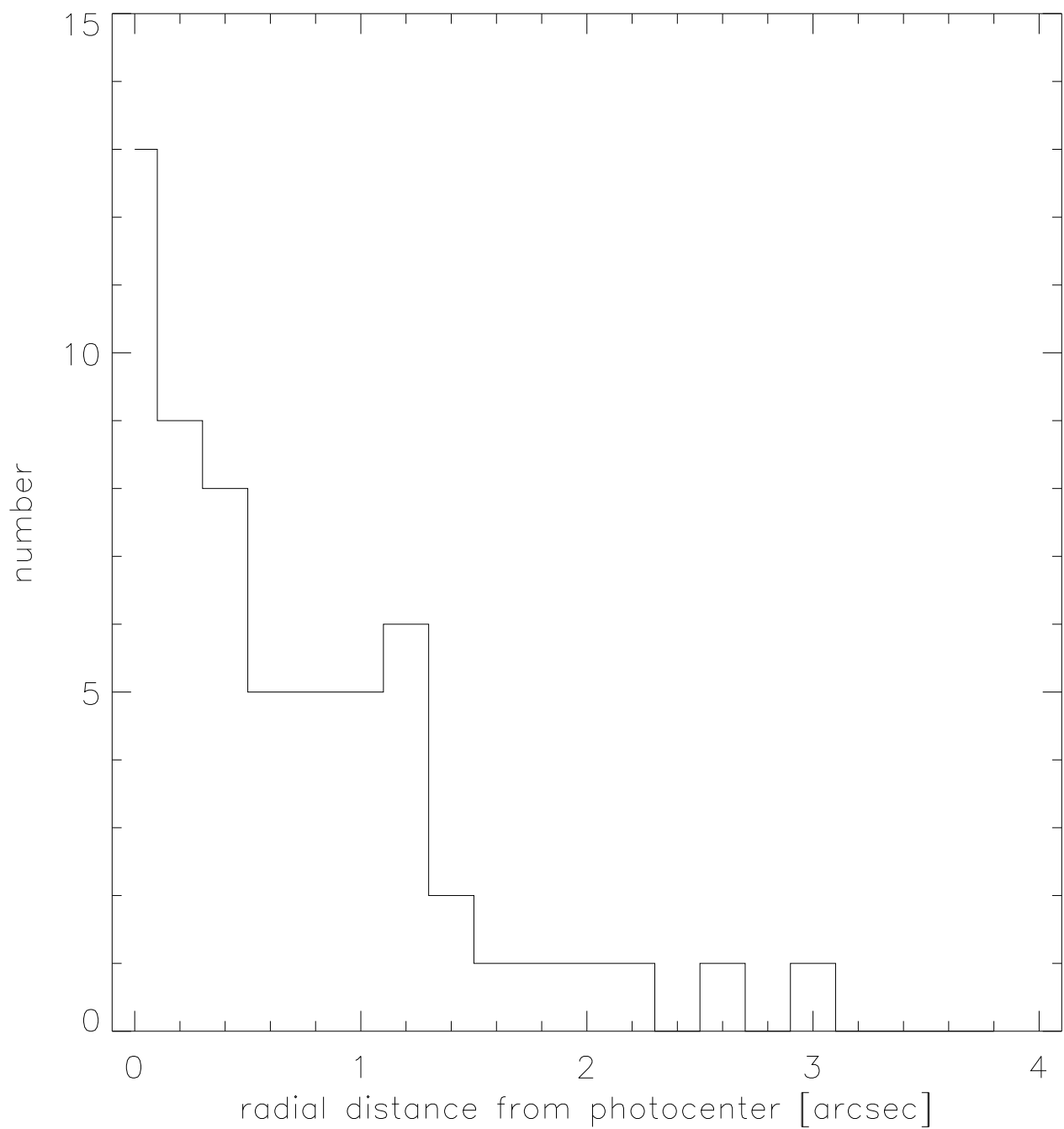
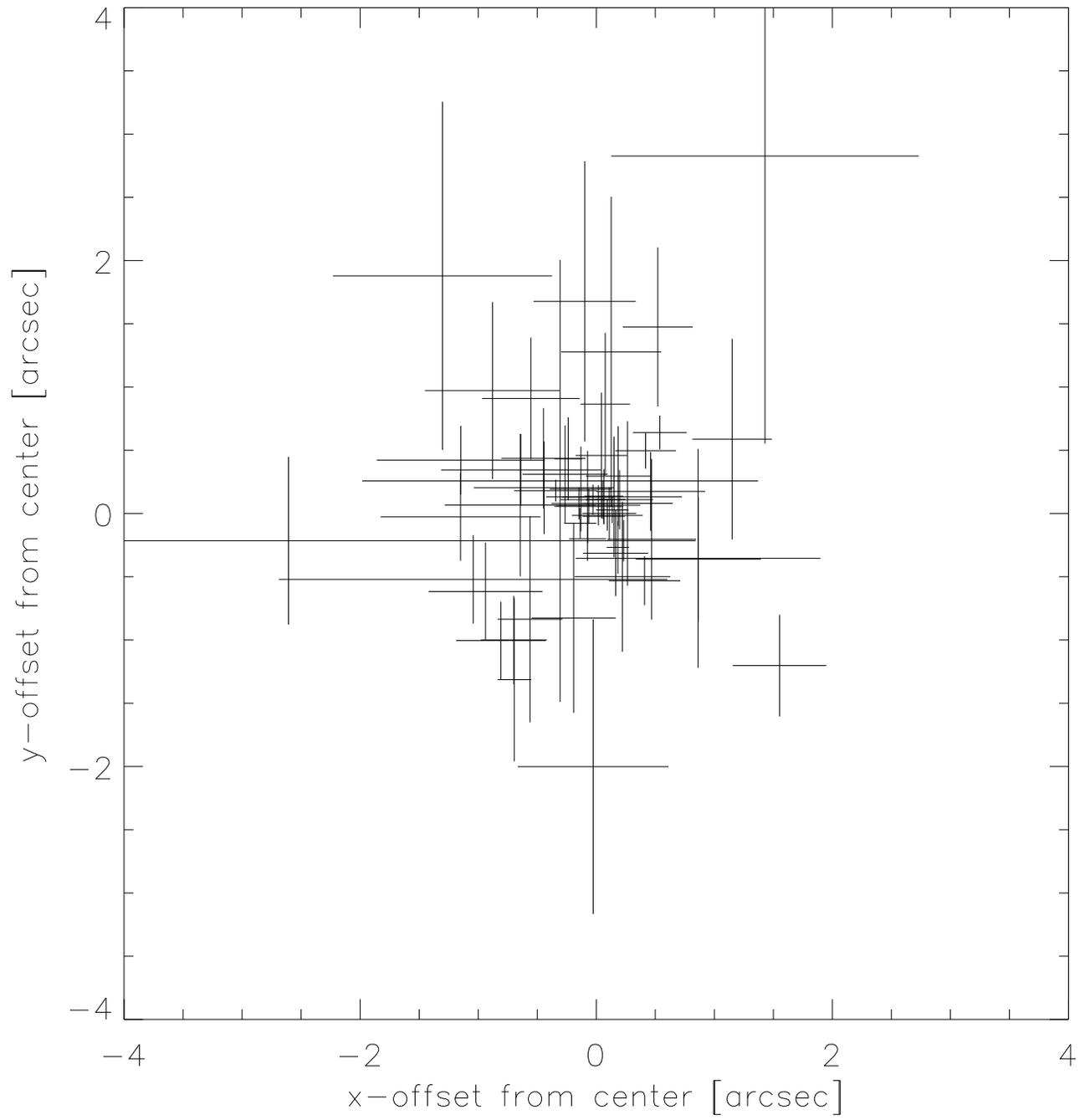
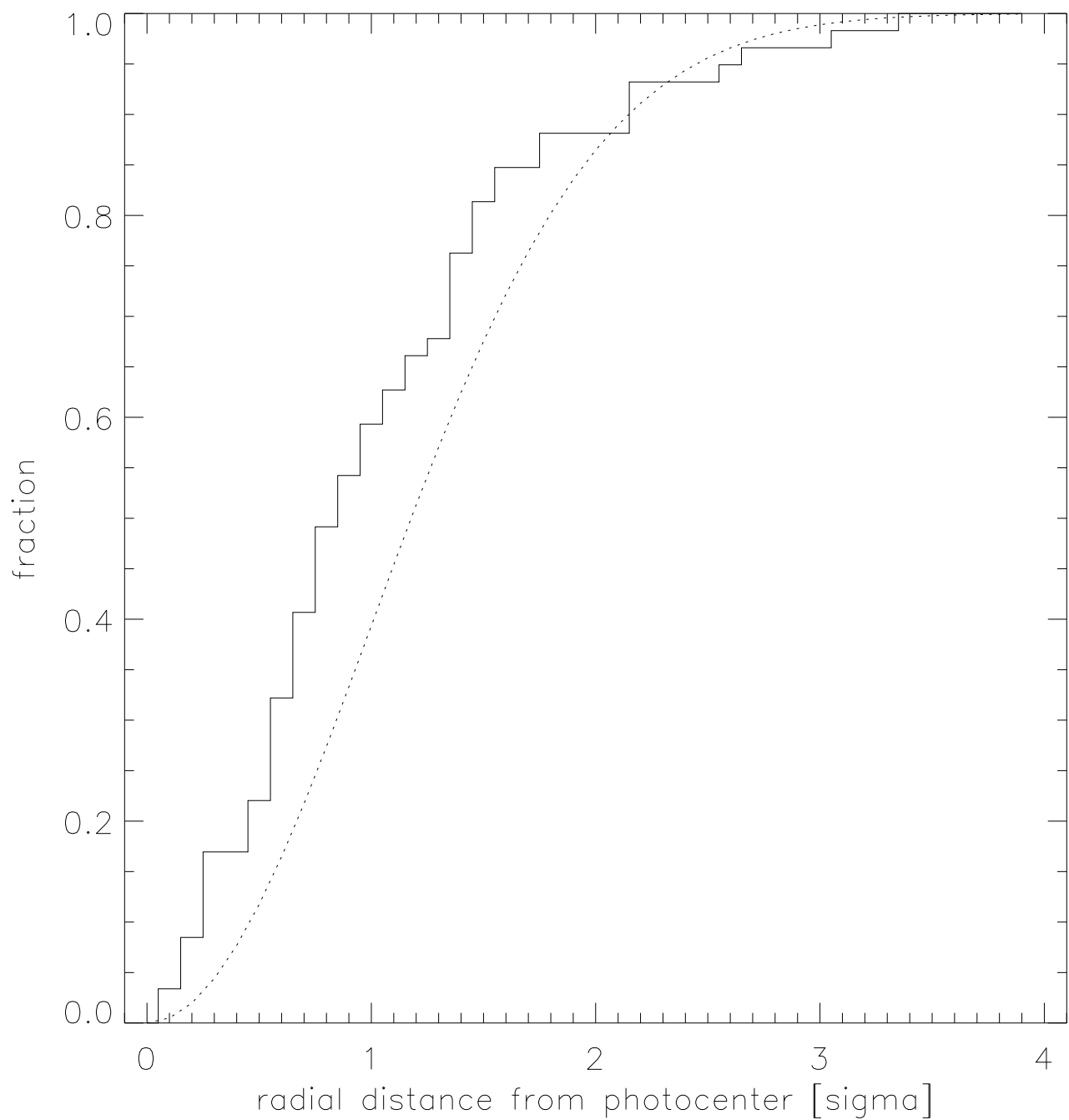
\end{document}